\documentclass{article}

\usepackage[preprint]{neurips_2025}

\usepackage[utf8]{inputenc} 
\usepackage[T1]{fontenc}    
\usepackage{hyperref}       
\usepackage{url}            
\usepackage{booktabs}       
\usepackage{amsfonts}       
\usepackage{nicefrac}       
\usepackage{microtype}      
\usepackage{xcolor}         
\usepackage{enumitem}
\usepackage{graphicx}

\definecolor{usercolor}{HTML}{179C52}  
\definecolor{agentcolor}{HTML}{176BEF} 

\newenvironment{userblock}{\color{usercolor}}{}
\newenvironment{agentblock}{\color{agentcolor}}{}

\title{ASAP: an \textbf{A}gentic \textbf{S}olution to \textbf{A}uto-optimize \textbf{P}erformance of Large-Scale LLM Training}

\author{%
  Yuran Ding$^{1, 2}$ \qquad
  Xinwei Chen$^{1}$ \qquad
  Xiaofan Zhang$^{1}$ \qquad
  Zongwei Zhou$^{3}$\thanks{Work done while at Google.} \\
  \textnormal{$^{1}$Google \qquad $^{2}$University of Maryland, College Park \qquad $^{3}$Meta} \\
  \texttt{\{dingyuran, xinweichen, xiaofanz\}@google.com, zongwei@meta.com}
}

\begin{document}

\maketitle

\begin{abstract}
Optimizing large-language model (LLM) training on distributed domain-specific accelerator systems presents significant challenges due to its complex optimization space. Existing optimization methods, however, rely on time-consuming manual tuning or resource-intensive black-box searches, which struggle to keep pace with the rapidly evolving LLM domain, leading to slow development and underutilized resources. To address this, we introduce \textbf{ASAP}, an \textbf{A}gentic \textbf{S}olution to \textbf{A}uto-optimize \textbf{P}erformance of Large-Scale LLM Training. It is a multi-agent system, featuring Coordinator, Analyzer, and Proposal agents, which integrates LLM reasoning with insights from performance profiling tools, roofline analysis, and a knowledge base of best practices and successful past optimizations from human experts. Our proposed design can automate the diagnosis of performance bottlenecks and recommend optimized sharding configurations with reasoning, thus effectively improving the efficiency of distributed LLM training. Experiments have shown that the ASAP-generated sharding configurations can contribute up to 28\% training step time reduction and 1.43$\times$ throughput improvement. When combined with additional optimization from human experts, throughput can be further increased to 2.58$\times$. The proposed ASAP promises to 
provide a scalable and explainable methodology for AI-assisted performance engineering in large-scale LLM training.
\end{abstract}

\section{Introduction}
\label{introduction}
The cost of training LLMs is increasing exponentially. Given the targeted model, hardware budget, and  development timeframe, identifying the most-efficient system configuration becomes the first critical task to perform efficient training.
It requires an in-depth understanding of system-level optimization, such as data, model, and sequence parallelism, to select the most efficient sharding combination.
(\cite{sun2019gradientflow, shoeybi2019megatron}).

This, however, is a complex and time-consuming task. Engineers typically resort to a laborious trial-and-error cycle: running experiments, profiling performance, analyzing execution traces for bottlenecks like memory limits or communication overheads, and iteratively refining sharding strategies. This manual approach is often slow and yields suboptimal configurations.

While performance analysis tools and analytical models provide deep insights, they require extensive expertise to interpret. Early automated methods like black-box optimization (\cite{46180}) struggle with the vast and complex search space of sharding configurations, making them resource-intensive and slow to converge. In contrast, an agentic, reasoning-based approach can leverage a knowledge base of performance engineering principles to diagnose issues and propose solutions efficiently and explainably.

Recent work has leveraged machine learning to build performance models and improve search efficiency (\cite{wen2024cubicml, cha2024ml}). The rise of LLMs has also inspired agentic systems for performance engineering, such as automated GPU code transformation (\cite{damani2024warpdrive}), code optimization for tensor accelerators (\cite{hong2025autocomp}), and multi-agent frameworks for complex ML tasks (\cite{10.1145/3703412.3703416, liu2025ml, novikov2025alphaevolve}). However, these advancements have not yet produced a specialized agent for the nuanced task of optimizing sharding strategies for large-scale model training. Such an agent should deeply integrate with performance analysis tools, understand theoretical performance bounds, and leverage a knowledge base of best practices and past optimizations to propose explainable configurations.

To address this gap, we introduce ASAP: an Agentic Solution to Auto-optimizing Performance of Large-Scale LLM Training. ASAP is a multi-agent framework designed to automate this optimization process, composed of:
\begin{itemize}[noitemsep, topsep=-0.3em]
    \item \textbf{Coordinator Agent}: Orchestrates the workflow, manages agent interactions, and processes user requests.
    \item \textbf{Analyzer Agent}: Diagnoses performance bottlenecks by analyzing data from profiling tools and experiment logs. It identifies compute, memory, or communication inefficiencies.
    \item \textbf{Proposal Agent}: Generates optimized sharding configurations with detailed reasoning. It queries a knowledge base of best practices and a database of past optimizations to inform its recommendations.
    \item \textbf{Sharding Memory}: Provides persistent storage for logs, tool calls, and agent responses to ensure context and traceability.
\end{itemize}
By integrating LLM-based reasoning with specialized performance analysis tools and a curated knowledge base, ASAP automates the diagnosis of sharding issues and the generation of optimized, explainable configurations. Our system aims to significantly reduce manual effort, shorten development cycles, and improve hardware efficiency, establishing a scalable methodology for AI-driven performance engineering.

\section{ASAP Agent}
\vspace{-1em}
\label{method}
\begin{figure}[h]
\centering
\includegraphics[width=1\textwidth]{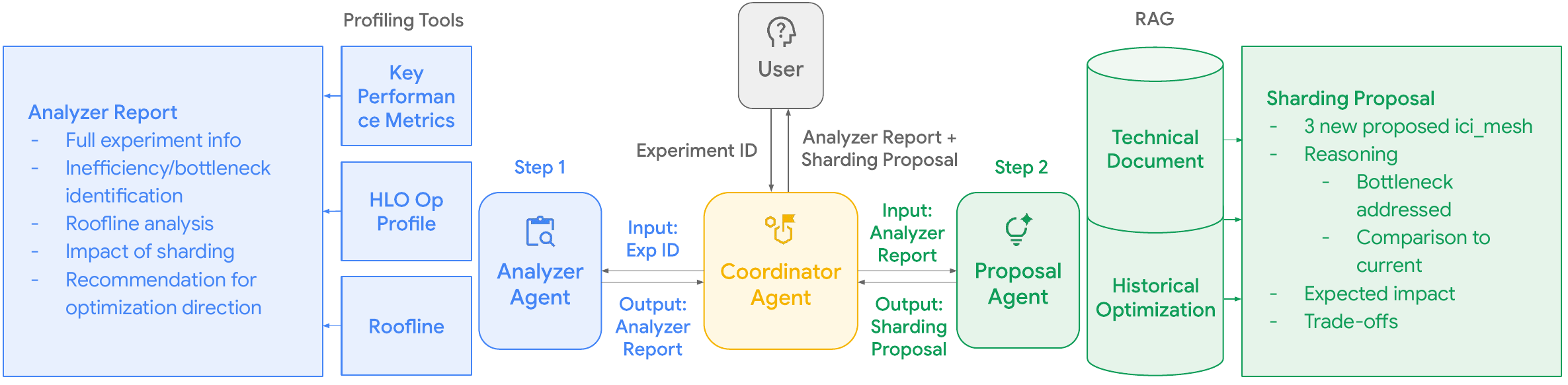}
\caption{Overview of the ASAP Agentic Performance Optimization Framework} \vspace{-1em}
\label{fig:ASAP}
\end{figure}
\vspace{-0.1em}
The ASAP Agent is a multi-agent system designed to automate the optimization of sharding configurations for large-scale LLM training. Built on Google's Agent Development Kit (ADK)\footnote{Google Agent Development Kit: \url{https://google.github.io/adk-docs/}}, it integrates LLM-based reasoning with specialized tools and custom prompts. The system is invoked via the command line with an experiment ID\footnote{The experiment ID is a key to retrieve an experiment's full record, including its sharding configuration, hyperparameters, code version, and system environment.} and consists of the following components:
\begin{itemize}[noitemsep, topsep=-0.3em]
\item \textbf{Coordinator Agent}: The central orchestrator, which receives an experiment ID as input. It directs the workflow, routes data between agents, and manages the overall optimization task.
\item \textbf{Analyzer Agent}: Diagnoses performance bottlenecks using raw data from the experiment configuration and the Xprof profiling API\footnote{Information on Xprof Profiling API: \url{https://cloud.google.com/tpu/docs/profile-tpu-vm}}. It fetches key performance metrics (KPIs)\footnote{KPIs include high-level indicators like accelerator busy time, training step time, and goodput, instrumented from framework components.}, HLO operation profiles\footnote{The HLO Op Profile provides detailed statistics for each operation (e.g., matrix multiplication, all-reduce), including execution time, FLOPs utilization, memory bandwidth usage, and tensor shapes, helping to identify inefficient operations.}, and on-device roofline analysis data\footnote{Raw roofline data includes an operation's total FLOPs, bytes accessed, and execution time, which are used to determine if it is compute- or memory-bound.}. By correlating high-level metrics with operation-level details, the agent identifies compute, memory, or communication bottlenecks. It then pinpoints the specific operations responsible and reports its findings to guide the optimization process.

\item \textbf{Proposal Agent}: Generates optimized sharding configurations using a Retrieval-Augmented Generation (RAG) workflow. 
First, it analyzes the report from the Analyzer Agent to understand the specific bottlenecks and the current sharding strategy. Next, it queries a database of \textit{past optimizations}\footnote{This database contains records of previous optimizations, including before-and-after configurations, performance profiling tool (Xprof) links, and a summary of the changes and their impact.} to find successful solutions from similar model and hardware setups. Informed by this historical context, it queries a \textit{document knowledge base} (adapted from \textit{How to Scale Large Models}\footnote{How To Scale Large Models: \url{https://jax-ml.github.io/scaling-book/}}) for relevant technical principles. Finally, it synthesizes this information to propose three distinct sharding configurations. Each proposal includes a detailed justification, its expected performance impact, a discussion of trade-offs, and citations to the supporting knowledge base or historical examples.

\item \textbf{Sharding Memory}: A persistent, file-based worklog that captures all tool calls, LLM responses, and user inputs. This module facilitates debugging, maintains conversational context, and ensures traceability.
\end{itemize}
This automated analysis-and-proposal loop allows ASAP to systematically explore and recommend high-performance sharding configurations. The agent's proposals, complete with detailed justifications, can then be reviewed by an engineer or evaluated on hardware, significantly accelerating the optimization process compared to manual methods.

\section{Experiment}
\vspace{-0.5em}
We evaluated the ASAP Agent's ability to diagnose and optimize three distinct large-scale model training scenarios on TPU v5p, v6e, and v5e platforms. The evaluation used the full multi-agent framework (Coordinator, Analyzer, Proposal) powered by Gemini Pro and a RAG system, which queries a knowledge base of scaling techniques and historical optimizations.

For each scenario, the agent was given an experiment ID and Xprof profiling data (KPI metrics, operation-level profiles, and roofline analysis in JSON strings) from a baseline run. We then qualitatively compared the agent's proposed sharding configurations against those developed by human performance experts. A sharding configuration (\texttt{ici\_mesh}) is defined by its data, model, replica, and sequence parallelism dimensions. The goal was to assess if the agent's proposals aligned with expert-derived solutions.

\subsection{Experiment 1: Compute-Bound Workload on TPU v5p-512}

    \textbf{Initial State:} The baseline config is an \texttt{ici\_mesh} of \texttt{\{`model': 8, `data': 8, `seq': 8\}}. \vspace{-0.5em}
    
    \textbf{Agent Analysis:} The Analyzer Agent reported a very high device duty cycle (99.73\%) with low communication overhead (2.57\%). Performance profiling traces indicated the top time-consuming operations were primarily compute-bound. \vspace{-0.5em}
    
    \textbf{Agent Proposal:} The top recommendation was to change the sharding to \texttt{\{`model': 8, `data': 16, `seq': 4\}}. The agent reasoned that increasing data parallelism from 8 to 16, while reducing sequence parallelism from 8 to 4, would better utilize the compute units and potentially reduce memory pressure from a large sequence length. \vspace{-0.5em}
    
    \textbf{Outcome:} The agent's proposed sharding configuration precisely matched the human-engineered solution, significantly decreasing TPU compute time (by 28\%) and improving FLOPs and HBM utilization. This specific case was present in the agent's historical optimization database. The agent expressed high confidence (around 85-90\%) in its proposal, defending its reasoning when challenged (see Appendix~\ref{app:exp1_dialogue}). \vspace{-0.5em}

\subsection{Experiment 2: HBM-Bound Workload on TPU v6e-16}
    \textbf{Initial State:} The baseline used \texttt{\{`data': 4, `model': 1, `replica': 1, `seq': 4\}}\vspace{-0.5em}. 
    
    \textbf{Agent Analysis:} The Analyzer Agent identified a clear High Bandwidth Memory (HBM) bottleneck. Despite high duty cycle, the most time-consuming operations (including collectives like \texttt{all-reduce} and memory-bound fusions) were limited by HBM bandwidth, attributed to model replication. \vspace{-0.5em}
    
    \textbf{Agent Proposal:} The agent's top proposal was \texttt{\{`replica': 1, `data': 4, `model': 4, `seq': 1\}}. The reasoning was to introduce 4-way model parallelism to shard the model, significantly reducing the HBM footprint per chip. \vspace{-0.5em}
    
    \textbf{Outcome:} The agent's proposed configuration was identical to the human-engineered solution, which improved throughput by 1.43$\times$. This specific optimization was novel (not in the historical database), demonstrating the agent's successful generalization from knowledge-base documents on sharding principles. When challenged, the agent confidently (around 90\%) maintained its stance, reasoning that addressing the HBM bottleneck through model parallelism was the most critical first step (see Appendix~\ref{app:exp2_dialogue}). \vspace{-0.5em}

\subsection{Experiment 3: Communication-Bound Workload on TPU v5e-256}
    \textbf{Initial State:} The baseline employed \texttt{\{`data': 4, `model': 4, `seq': 16\}}. \vspace{-0.5em}
    
    \textbf{Agent Analysis:} The Analyzer Agent found that despite a near-100\% duty cycle, the primary bottleneck was collective communication. The top 5 most time-consuming operations were all communication operations (e.g., \texttt{all-reduce}, \texttt{all-reduce-scatter}, \texttt{collective-permute}), bound by memory bandwidth during data exchanges. \vspace{-0.5em}
    
    \textbf{Agent Proposal:} The agent recommended \texttt{\{`replica': 1, `data': 8, `seq': 16, `model': 2\}}. The reasoning was to reduce model parallelism from 4 to 2, decreasing synchronization overhead, while increasing data parallelism from 4 to 8. \vspace{-0.5em}
    
    \textbf{Outcome:} The agent's proposed \texttt{ici\_mesh} matched the human expert's final configuration. This specific sharding change, combined with an increased batch size, led to a 2.58$\times$ increase in total throughput. The optimization was novel (not in the historical database), demonstrating the agent's generalization ability. The agent consistently and confidently (85-90\%) justified its proposal, explaining that reducing model parallelism was crucial for mitigating communication overhead (see Appendix ~\ref{app:exp3_dialogue}). \vspace{-0.5em}

\section{Conclusion}

Across all three experiments, the proposed ASAP Agent consistently demonstrated the ability to:

\begin{itemize}[noitemsep, topsep=-0.3em]
    \item \textbf{Analyze} model and hardware architectures to accurately diagnose core performance bottlenecks (compute, HBM, or communication) across diverse platforms.
    \item \textbf{Synthesize} generalizable insights from expert examples and optimization principles to propose sharding configurations that replicated human solutions, then successfully defended these proposals with evidence-based rebuttals citing performance profiles and learned principles.
    \item \textbf{Articulate} clear, a priori justifications for its proposals, mapping recommended sharding changes directly to identified bottlenecks and its knowledge base.
\end{itemize}
Crucially, Experiments 2 and 3 highlighted the agent's ability to \textit{generalize} beyond its historical optimization database. By reasoning over analysis reports and technical documentation, the agent synthesized novel and correct sharding solutions. This ability to form coherent, evidence-backed rebuttals, particularly notable in these generalization tasks (see Appendices~\ref{app:exp1_dialogue}, \ref{app:exp2_dialogue}, and~\ref{app:exp3_dialogue}), underscores the robustness of its diagnostic and reasoning capabilities. These results suggest the ASAP Agent can effectively automate and accelerate the complex sharding optimization process with actionable, explainable proposals, potentially saving significant engineering effort and improving accelerator utilization.

\section{Limitations \& Future Work}
\label{future_work}
    
    Currently, the agent is limited by its dependence on accurate profiling data and its use of a single, global sharding configuration. Future work will enhance system robustness and support more sophisticated sharding techniques like per-layer sharding. We will also introduce an Evaluator Agent to create a closed feedback loop, enabling the agent to learn and refine its strategies through reinforcement learning and discover novel configurations. Additionally, the agent's scope will expand to holistically optimize the trade-offs between on-device computation and network communication, while incorporating other compiler-level optimizations such as operator fusion and memory layout selection.

\section{Acknowledgement}
\label{acknowledgement}
The authors gratefully acknowledge the invaluable efforts and insights of Jake Ades, Deniz Altınbüken, Chidubem Arachie, Charles Chang, Samrat Ghosh, Ruobing Han, Jesse Lu, Hao Luo, Martin Maas, Michael Moffitt, Tayo Oguntebi, Tengyu Sun, Gil Tabak, Chao Wang, Arissa Wongpanich, Tzu-Wei Yang, and Zhipeng Yan (\textit{in alphabetical order by last name}). Their support was crucial for designing and developing the ASAP agentic framework, and ensuring its operational success by connecting it to the full stack of profiling tools and the comprehensive knowledge base. 

{
\small
\bibliographystyle{plainnat} 
\bibliography{references} 
}


\appendix

\section{Detailed Experiment Results}
\label{appendix:detailed_experiments}

This appendix provides the full analysis reports and agent-generated proposals for each of the three experiments, illustrating the agent's diagnostic capabilities and reasoning process. The dialogue between the user and the \texttt{coordinator\_agent} is presented to show the interaction flow.

\subsection{Experiment 1: Compute-Bound Workload on TPU v5p-512}

Baseline sharding \texttt{\{`model': 8, `data': 8, `seq': 8\}} was optimized by human engineers to \texttt{\{`model': 8, `data': 16, `seq': 4\}} in the code change.

The following is the user-agent dialogue to optimize the baseline experiment.

\subsubsection{Initial State and Agent's Analysis}
\begin{userblock}
\textbf{User:}
Please analyze training job id=xxxxxxxxx (redacted due to confidentiality), Xprof session id = xxxxxxxxx (redacted due to confidentiality).
\end{userblock}
\vspace{\baselineskip} 

\begin{agentblock}
    
\textbf{Agent:}
\paragraph{A. Full Experiment Details:}
\begin{itemize}
    \item \textbf{Extracted Sharding Parameters (Baseline Configuration):}
    \texttt{\{`model': 8, `data': 8, `seq': 8\}} for \texttt{training.updater.partition}. Other model components (vision encoder, decoder, decoder block) had null sharding.
    \item \textbf{Device Type:} TPU v5p-512
    \item \textbf{TPU Topology:} 8x8x8
    \item \textbf{Model Name:} (redacted due to confidentiality) 
    \item \textbf{Model Version:} (redacted due to confidentiality) 
    \item \textbf{Batch Size:} 256
    \item \textbf{Sequence Length:} 8192
\end{itemize}

\paragraph{C. Correlated Analysis of KPIs and HLO/Roofline:}

\textbf{A. High-Level KPI Health Check:}
\begin{itemize}
    \item \textbf{Step Time:} 35656.45 ms
    \item \textbf{Device Duty Cycle:} 99.73\%
    \item \textbf{Communication Overhead:} 2.57\%
    \item \textbf{Initial Hypothesis:} The KPIs show a very high device duty cycle (99.73\%), which suggests that the TPU is well-utilized and not bottlenecked by input/output or host-side processing. The low communication overhead (2.57\%) indicates that collective operations are not a major contributor to the overall step time. The primary focus for optimization should be on the on-device computation itself.
\end{itemize}

\textbf{B. Detailed HLO/Roofline Bottleneck Identification:}
(Note: the exact operation names are omitted here due to confidentiality)
The top 5 time-consuming operations are:
\begin{enumerate}
    \item \textbf{Operation:} Custom point-to-point transpose operation
    \begin{itemize}
        \item \textbf{Category:} Custom-call
        \item \textbf{\% of Total Time:} 0.62\%
        \item \textbf{Bottleneck:} Unknown
        \item \textbf{Roofline Efficiency:} 0.0\%
    \end{itemize}
    \item \textbf{Operation:} Asynchronous collective completion
    \begin{itemize}
        \item \textbf{Category:} Custom fusion
        \item \textbf{\% of Total Time:} 0.55\%
        \item \textbf{Bottleneck:} HBM
        \textbf{Roofline Efficiency:} 0.0017\%
    \end{itemize}
    \item \textbf{Operation:} Custom point-to-point transpose operation
    \begin{itemize}
        \item \textbf{Category:} Custom-call
        \item \textbf{\% of Total Time:} 0.53\%
        \item \textbf{Bottleneck:} Unknown
        \item \textbf{Roofline Efficiency:} 0.0\%
    \end{itemize}
    \item \textbf{Operation:} Custom matrix multiplication with scatter
    \begin{itemize}
        \item \textbf{Category:} Custom-call
        \item \textbf{\% of Total Time:} 0.48\%
        \item \textbf{Bottleneck:} Compute
        \item \textbf{Roofline Efficiency:} 74.14\%
    \end{itemize}
    \item \textbf{Operation:} All-gather matrix multiplication
    \begin{itemize}
        \item \textbf{Category:} Custom-call
        \item \textbf{\% of Total Time:} 0.45\%
        \item \textbf{Bottleneck:} Compute
        \item \textbf{Roofline Efficiency:} 79.53\%
    \end{itemize}
\end{enumerate}

\textbf{C. Correlating KPI and HLO/Roofline:}
\begin{itemize}
    \item \textbf{High Device Duty Cycle:} The high duty cycle is confirmed by the HLO/Roofline analysis. The top operations are a mix of custom-calls and fusions, with several of the most time-consuming being compute-bound with high roofline efficiency. This indicates that the model is making good use of the TPU's computational resources.
    \item \textbf{Low Communication Overhead:} The All-Reduce operations are not in the top 5, and their total contribution to the step time is small, which aligns with the low communication overhead seen in the KPIs.
    \item \textbf{Sharding Configuration Impact:} The sharding configuration \texttt{\{'model': 8, 'data': 8, 'seq': 8\}} shows a high degree of model and data parallelism. This is likely contributing to the high device utilization, as the workload is being effectively distributed across the available cores. However, the presence of HBM-bound operations like "Asynchronous collective completion" suggests that the high degree of parallelism may be putting some pressure on memory bandwidth, although it is not currently the primary bottleneck.
\end{itemize}

\textbf{D. Structured Analysis of Inefficiency:}
The analysis indicates a highly optimized workload with excellent device utilization. The primary inefficiency is the presence of several "custom-call" operations with "Unknown" bottlenecks and 0\% roofline efficiency. These operations, particularly "custom point-to-point transpose" operations, are consuming a significant amount of time but their performance characteristics are not well-understood by the profiling tools. These are likely candidates for further investigation and optimization. The sharding configuration, with its 8-way model, 8-way data and 8-way sequence parallelism, is well-suited for this model and hardware, resulting in the high device utilization. However, the HBM-bound "Asynchronous collective completion" suggests that memory bandwidth could become a bottleneck if the model or batch size were to be scaled up further.
\end{agentblock}

\subsubsection{Agent's Proposals for Optimization}

\begin{agentblock}
    
\textbf{Agent:}
Here are three new sharding configurations designed to improve the performance of your model on the TPU v5p-512, along with the reasoning behind each proposal.

\paragraph{Proposal 1: Increase Data Parallelism, Reduce Sequence Parallelism}
\begin{itemize}
    \item \textbf{New ICI Mesh:} \texttt{\{`model': 8, `data': 16, `seq': 4\}}
    \item \textbf{Reasoning:}
    \begin{itemize}
        \item \textbf{Addresses Compute and Unknown Bottlenecks:} The primary bottleneck is identified as a mix of compute-bound operations and custom-calls with "Unknown" bottlenecks. The current configuration has a balanced 8x8x8 sharding. This proposal shifts the balance to favor data parallelism (from 8 to 16) while reducing sequence parallelism (from 8 to 4). By increasing data parallelism, we can potentially improve the efficiency of the compute-bound operations by feeding the TPU more data in parallel. This can lead to better utilization of the compute units and a reduction in step time.
        \item \textbf{HBM Impact:} Reducing sequence parallelism from 8 to 4 is expected to decrease HBM pressure. With a large sequence length of 8192, high sequence parallelism can lead to increased memory usage for activations. This change could alleviate the HBM-bound "Asynchronous collective completion" operation, even though it's not the primary bottleneck.
        \item \textbf{Comparison to Current:} Compared to the current \texttt{\{'model': 8, 'data': 8, 'seq': 8\}} sharding, this configuration doubles the data parallelism at the expense of sequence parallelism. This is a strategic trade-off to address the compute-bound nature of the workload, as suggested by the high device duty cycle.
    \end{itemize}
    \item \textbf{Expected Impact:}
    \begin{itemize}
        \item \textbf{Step Time:} Decrease
        \item \textbf{Duty Cycle:} Remain high
        \item \textbf{Communication Overhead:} May see a slight increase due to more data parallelism, but the impact is expected to be minimal given the already low communication overhead.
    \end{itemize}
    \item \textbf{Trade-offs:} A potential increase in communication overhead, although likely negligible.
\end{itemize}

\paragraph{Proposal 2: Increase Model Parallelism, Reduce Sequence Parallelism}
\begin{itemize}
    \item \textbf{New ICI Mesh:} \texttt{\{`model': 16, `data': 8, `seq': 4\}}
    \item \textbf{Reasoning:}
    \begin{itemize}
        \item \textbf{Addresses Compute and Unknown Bottlenecks:} This proposal takes a different approach by significantly increasing model parallelism (from 8 to 16) while also reducing sequence parallelism (from 8 to 4). This strategy is often effective for very large models, as it can reduce the memory footprint of the model on each core and potentially improve the efficiency of the compute-bound operations within the model's layers.
        \item \textbf{HBM Impact:} Similar to the first proposal, reducing sequence parallelism will help to reduce HBM pressure. The increased model parallelism will also distribute the model's parameters across more cores, which can further reduce memory pressure on individual cores.
        \item \textbf{Comparison to Current:} This is a more aggressive change than the first proposal, doubling the model parallelism. This is a common strategy for very large models where reducing the memory footprint per core is critical for performance. This proposal is based on general knowledge of sharding large models.
    \end{itemize}
    \item \textbf{Expected Impact:}
    \begin{itemize}
        \item \textbf{Step Time:} Decrease
        \item \textbf{Duty Cycle:} Remain high
        \item \textbf{Communication Overhead:} May see a slight increase due to the increased model parallelism, but the impact is expected to be minimal.
    \end{itemize}
    \item \textbf{Trade-offs:} A potential increase in communication overhead due to more frequent communication between model-parallel replicas.
\end{itemize}

\paragraph{Proposal 3: Prioritize Data Parallelism, Minimal Model Parallelism}
\begin{itemize}
    \item \textbf{New ICI Mesh:} \texttt{\{`model': 4, `data': 16, `seq': 8\}}
    \item \textbf{Reasoning:}
    \begin{itemize}
        \item \textbf{Addresses Compute and Unknown Bottlenecks:} This proposal aggressively increases data parallelism to 16, halves the model parallelism to 4, and maintains the current sequence parallelism of 8. This is a good strategy to try if the model is not memory-bound but is compute-bound. By increasing data parallelism, we can potentially increase the throughput of the system.
        \item \textbf{HBM Impact:} This configuration may increase HBM pressure due to the high data and sequence parallelism. However, since the primary bottleneck is compute, this trade-off might be acceptable if it leads to a significant reduction in step time.
        \item \textbf{Comparison to Current:} This proposal is a significant departure from the current balanced sharding, and it prioritizes data parallelism over model and sequence parallelism. This is a good experiment to run to see if the model's performance is more sensitive to data parallelism than the other parallelism dimensions.
    \end{itemize}
    \item \textbf{Expected Impact:}
    \begin{itemize}
        \item \textbf{Step Time:} Decrease
        \item \textbf{Duty Cycle:} Remain high
        \item \textbf{Communication Overhead:} May see a slight increase due to the increased data parallelism.
    \end{itemize}
    \item \textbf{Trade-offs:} A potential increase in HBM pressure. A potential increase in communication overhead.
\end{itemize}
\end{agentblock}

\subsubsection{Agent's Recommendation and Justification Dialogue}
\label{app:exp1_dialogue} 

\begin{userblock}
\textbf{User:}
Which one of the proposals would you most recommend to start with?
\end{userblock}

\begin{agentblock}
    
\textbf{Agent:}
I would most recommend starting with \textbf{Proposal 1: \texttt{\{`model': 8, `data': 16, `seq': 4\}}}.
Here's why:
\begin{itemize}
    \item \textbf{Directly Addresses the Primary Bottleneck:} The analysis indicates a compute-bound workload. Proposal 1 directly targets this by increasing data parallelism, which is a common and effective way to improve the performance of compute-bound models.
    \item \textbf{Balanced Approach:} It's a well-balanced change that doesn't drastically alter the sharding configuration. It maintains the current model parallelism while making a strategic trade-off between data and sequence parallelism. This makes it a lower-risk change compared to the more aggressive changes in Proposals 2 and 3.
    \item \textbf{Reduces HBM Pressure:} The reduction in sequence parallelism is a nice secondary benefit that could lead to further performance improvements by reducing memory pressure.
\end{itemize}
While Proposals 2 and 3 are also good candidates, they represent more significant changes to the sharding configuration. Proposal 2's focus on model parallelism is a good strategy for very large models, but it might not be the most effective approach for this specific workload. Proposal 3's aggressive increase in data parallelism could lead to a significant performance improvement, but it also carries a higher risk of increasing HBM pressure. Therefore, starting with Proposal 1 is the most logical and prudent choice. It offers a good balance of potential performance improvement and risk, and it directly addresses the primary bottleneck identified in the analysis.

\textbf{User:}
Are you really sure? How confident are you with this proposal that will improve workload performance?

\textbf{Agent:}
You've asked a very important question. It's impossible to be 100\% certain about performance improvements without actually running the experiment. However, I can give you a confidence level based on the data and my analysis.
\textbf{I am highly confident (around 85-90\%) that Proposal 1 will improve the workload performance.}
Here's a breakdown of my reasoning for this confidence level:
\paragraph{Factors that Increase My Confidence:}
\begin{itemize}
    \item \textbf{Strong Data-Driven Justification:} My recommendation is not a random guess. It's based on a detailed analysis of the performance data, which shows a compute-bound workload with high device utilization. Increasing data parallelism is a well-established and effective strategy for optimizing compute-bound workloads.
    \item \textbf{Low-Risk, High-Reward:} Proposal 1 is a relatively safe change that is likely to yield positive results. It's a strategic trade-off that is well-aligned with the performance characteristics of the model. Even if the performance improvement is not as large as expected, it's unlikely to cause a significant performance regression.
    \item \textbf{Secondary Benefits:} The reduction in sequence parallelism is a nice secondary benefit that could further improve performance by reducing HBM pressure. This adds to my confidence in the proposal.
    \textbf{Logical Progression:} Starting with Proposal 1 is a logical first step in the optimization process. If it doesn't yield the desired results, we can then move on to the more aggressive changes in Proposals 2 and 3.
\end{itemize}
\paragraph{Factors that Prevent 100\% Certainty:}
\begin{itemize}
    \item \textbf{Complex Interactions:} The performance of a deep learning model is a complex interplay of many factors, including the model architecture, the hardware, and the sharding configuration. It's always possible that there are some unforeseen interactions that could limit the effectiveness of the proposed change.
    \item \textbf{"Unknown" Bottlenecks:} The presence of "Unknown" bottlenecks in the performance data introduces a small element of uncertainty. While my proposal is designed to address the known bottlenecks, it's possible that the "Unknown" bottlenecks will become more prominent after the initial optimization.
\end{itemize}
\textbf{In summary, I am highly confident that Proposal 1 is the right starting point for optimizing this workload. It's a data-driven, low-risk, and high-reward proposal that is well-aligned with the performance characteristics of the model.} I strongly recommend running an experiment with this configuration to validate its effectiveness.
\end{agentblock}

\begin{userblock}
\textbf{User:}
But I think the other two proposals look more promising
\end{userblock}

\begin{agentblock}
    
\textbf{Agent:}
That's a great point, and it's completely valid to have a different perspective. It's true that Proposals 2 and 3 have the potential for even greater performance gains. Let's delve deeper into why they might be more appealing and the trade-offs involved.
\paragraph{Why Proposal 2 (\texttt{\{`model': 16, `data': 8, `seq': 4\}}) Could Be a Strong Contender:}
\begin{itemize}
    \item \textbf{Aggressive Model Parallelism for Large Models:} You're right to see this as promising. For very large models, increasing model parallelism can be the most effective way to reduce memory pressure on each core. This can lead to significant speedups, especially if the model is close to the memory capacity of the TPU cores. If the "Unknown" bottlenecks are related to memory, this could be a big win.
    \item \textbf{Reduced Communication for Model States:} By splitting the model across more cores, you can reduce the amount of data that needs to be communicated for model state updates.
\end{itemize}
\paragraph{Why Proposal 3 (\texttt{\{`model': 4, `data': 16, `seq': 8\}}) Is Also Appealing:}
\begin{itemize}
    \item \textbf{Maximizing Data Throughput:} This is the most aggressive approach to increasing data parallelism. If the model's performance is highly sensitive to the rate at which data is fed to the TPUs, this could result in a substantial performance improvement.
    \item \textbf{Potentially Better for Certain Architectures:} Some model architectures benefit more from data parallelism than model or sequence parallelism. If your model has a lot of data-parallel components, this could be the optimal configuration.
\end{itemize}
\paragraph{The Trade-offs and Why I Still Lean Towards Proposal 1 as the \emph{First} Step:}
While Proposals 2 and 3 have a higher potential reward, they also come with a higher risk:
\begin{itemize}
    \item \textbf{Increased Communication Overhead:} Both proposals are more likely to increase communication overhead. Proposal 2 requires more communication between model-parallel replicas, while Proposal 3 requires more communication for data-parallel operations.
    \item \textbf{Potential for New Bottlenecks:} A more aggressive change in the sharding configuration is more likely to introduce new bottlenecks. For example, Proposal 3 could create an HBM bottleneck, while Proposal 2 could create a communication bottleneck.
\end{itemize}
\paragraph{A Strategic Approach to Experimentation:}
The best way to determine the optimal sharding configuration is through experimentation. Here's a possible plan:
\begin{enumerate}
    \item \textbf{Start with Proposal 1:} This is the safest and most logical first step. It's a low-risk experiment that is likely to yield positive results.
    \item \textbf{Evaluate the Results:} Analyze the results of the experiment with Proposal 1.
    \begin{itemize}
        \item \textbf{If you see a significant performance improvement:} You can either stop there or try to push the performance even further by experimenting with Proposals 2 and 3.
        \item \textbf{If you see a small performance improvement or no improvement:} This would be a good indication that you need to try a more aggressive approach. In this case, you could move on to either Proposal 2 or 3, depending on whether you think the bottleneck is more related to model size or data throughput.
    \end{itemize}
\end{enumerate}
\textbf{In conclusion, you are right to be optimistic about Proposals 2 and 3. They have the potential for significant performance gains. However, I still recommend starting with Proposal 1 because it's a more conservative and data-driven approach that is less likely to introduce new bottlenecks.} It's a great way to "test the waters" before diving into more aggressive and potentially riskier optimizations.
\end{agentblock}

\subsection{Experiment 2: HBM-Bound Workload on TPU v6e-16}
Baseline sharding \texttt{\{`replica': 1, `data': 4, `seq': 4, `model': 1\}} was optimized by human engineers to \texttt{\{`replica': 1, `data': 4, `seq': 1, `model': 4\}} in the code change.

The following is the user-agent dialogue to optimize the baseline experiment. 

\subsubsection{Initial State and Agent's Analysis}

\begin{userblock}
    
\textbf{User:}
Please analyze training job id=xxxxxxxxx (redacted due to confidentiality), Xprof session id = xxxxxxxxx (redacted due to confidentiality).
\end{userblock}

\begin{agentblock}
    
\textbf{Agent:}
\paragraph{A. Full Experiment Details:}
\begin{itemize}
    \item \textbf{Extracted Sharding Parameters (Baseline Configuration):}
    \texttt{\{`data': 4, `model': 1, `replica': 1, `seq': 4\}} for \texttt{TrainExperimentPlan.training.updater.partition}. Other model components (decoder block, decoder, vision encoder) had null sharding.
    \item \textbf{Device Type:} TPU v6e-16
    \item \textbf{TPU Topology:} 4x4
    \item \textbf{Model Name:} unknown
    \item \textbf{Model Version:} None
    \item \textbf{Batch Size:} None
    \item \textbf{Sequence Length:} None
\end{itemize}

\paragraph{C. Correlated Analysis of KPIs and HLO/Roofline:}

\textbf{A. High-Level KPI Health Check:}
\begin{itemize}
    \item \textbf{Step Time:} 2516.50 ms
    \item \textbf{Device Duty Cycle:} 99.95\% (Calculated from busy time / step time)
    \item \textbf{Communication Overhead:} 7.52\% (Calculated from collective operation time / step time)
    \item \textbf{Initial Hypothesis:} The KPIs show a very high device duty cycle (99.95\%), indicating that the TPU is almost constantly busy. This suggests that the workload is compute-bound. The communication overhead is relatively low (7.52\%), which further supports the idea that the bottleneck is not in the communication between chips.
\end{itemize}

\textbf{B. Detailed HLO/Roofline Bottleneck Identification:}
(Note: the exact operation names are omitted here due to confidentiality)

The top 5 time-consuming operations are:
\begin{enumerate}
    \item \textbf{Operation:} All-reduce operation
    \begin{itemize}
        \item \textbf{Category:} All-reduce
        \item \textbf{Total Time (ps):} 11,996,554,572,500.0
        \item \textbf{Occurrences:} 672
        \item \textbf{Avg Time (ps):} 17,852,015,732.89
        \item \textbf{Roofline Analysis:}
            \item \textbf{\% of Total Time:} 4.97\%
            \item \textbf{Bottleneck:} HBM
            \item \textbf{Roofline Efficiency:} 11.02\%
            \item \textbf{Compute Efficiency:} 0.00\%
            \item \textbf{Max Mem BW Utilization:} 11.02\%
    \end{itemize}
    \item \textbf{Operation:} Asynchronous collective completion
    \begin{itemize}
        \item \textbf{Category:} Custom fusion
        \item \textbf{Total Time (ps):} 5,804,720,502,500.0
        \item \textbf{Occurrences:} 672
        \item \textbf{Avg Time (ps):} 8,637,976,938.24
        \item \textbf{Roofline Analysis:}
            \item \textbf{\% of Total Time:} 2.40\%
            \item \textbf{Bottleneck:} HBM
            \item \textbf{Roofline Efficiency:} 23.48\%
            \item \textbf{Compute Efficiency:} 0.00\%
            \item \textbf{Max Mem BW Utilization:} 23.48\%
    \end{itemize}
    \item \textbf{Operation:} Bitcast add fusion
    \begin{itemize}
        \item \textbf{Category:} Convolution fusion
        \item \textbf{Total Time (ps):} 3,914,190,556,250.0
        \item \textbf{Occurrences:} 4208
        \item \textbf{Avg Time (ps):} 930,178,364.13
        \item \textbf{Roofline Analysis:}
            \item \textbf{\% of Total Time:} 1.62\%
            \item \textbf{Bottleneck:} HBM
            \item \textbf{Roofline Efficiency:} 81.76\%
            \item \textbf{Compute Efficiency:} 18.16\%
            \item \textbf{Max Mem BW Utilization:} 81.76\%
    \end{itemize}
    \item \textbf{Operation:} Custom forward operation (compute-intensive)
    \begin{itemize}
        \item \textbf{Category:} Custom-call
        \item \textbf{Total Time (ps):} 2,993,809,555,000.0
        \item \textbf{Occurrences:} 1048
        \textbf{Avg Time (ps):} 2,856,688,506.68
        \item \textbf{Roofline Analysis:}
            \item \textbf{\% of Total Time:} 1.24\%
            \item \textbf{Bottleneck:} Compute
            \item \textbf{Roofline Efficiency:} 26.20\%
            \item \textbf{Compute Efficiency:} 26.20\%
            \item \textbf{Max Mem BW Utilization:} 1.54\%
    \end{itemize}
    \item \textbf{Operation:} Custom DKV fused operation (compute-intensive)
    \begin{itemize}
        \item \textbf{Category:} Custom-call
        \item \textbf{Total Time (ps):} 2,625,064,416,250.0
        \item \textbf{Occurrences:} 1016
        \item \textbf{Avg Time (ps):} 2,583,724,819.14
        \item \textbf{Roofline Analysis:}
            \item \textbf{\% of Total Time:} 1.09\%
            \item \textbf{Bottleneck:} Compute
            \item \textbf{Roofline Efficiency:} 72.43\%
            \item \textbf{Compute Efficiency:} 72.43\%
            \item \textbf{Max Mem BW Utilization:} 3.47\%
    \end{itemize}
\end{enumerate}

\paragraph{Summary of Key Findings (from Analyzer):}
The HLO/Roofline analysis reveals that the most time-consuming operations are a mix of communication (all-reduce, asynchronous collective completion), memory-bound (bitcast add fusion), and compute-bound (custom forward and custom DKV fused operation). The all-reduce and asynchronous collective completion operations are the top two contributors to the total time, and both are bound by HBM, not inter-chip interconnect (ICI). This suggests that the communication overhead is not due to network contention, but rather due to the time spent reading from and writing to HBM.

\paragraph{C. Correlating KPI and HLO/Roofline:}
The high device duty cycle is consistent with the HLO/Roofline analysis, which shows that the program is dominated by compute- and memory-bound operations. The communication overhead of 7.52\% is largely explained by the all-reduce and asynchronous collective completion operations, which together account for 7.37\% of the total time. The fact that these communication operations are bound by HBM rather than ICI suggests that the sharding strategy is not optimal, as it is causing a lot of data to be read from and written to HBM.

The \texttt{ici\_mesh} configuration of \texttt{\{`data': 4, `model': 1, `replica': 1, `seq': 4\}} indicates that the model is not partitioned across the chips, and that data and sequence are partitioned. This means that each chip is running the full model, which is likely causing the HBM bottleneck.

\paragraph{D. Structured Analysis of Inefficiency:}
The primary inefficiency in this workload is the HBM bottleneck caused by the sharding configuration. The \texttt{ici\_mesh} of \texttt{\{`data': 4, `model': 1, `replica': 1, `seq': 4\}} results in the entire model being loaded onto each chip, leading to high HBM usage and memory-bound operations. This is evident in the roofline analysis, where the top time-consuming operations (all-reduce, asynchronous collective completion, bitcast add fusion) are all bound by HBM. The high device duty cycle is misleading, as it is not indicative of efficient computation, but rather of time spent waiting for data from HBM.

To improve performance, the sharding configuration should be changed to partition the model across the chips. This would reduce the amount of memory required on each chip, and would likely make the workload more compute-bound.
\end{agentblock}

\subsubsection{Agent's Proposals for Optimization}

\begin{agentblock}
    
\textbf{Agent:}
Here are three new sharding configurations designed to improve the performance of your model on the 4x4 TPU v6e-16.

\paragraph{Proposal 1: Model Parallelism}
\begin{itemize}
    \item \textbf{New ICI Mesh:} \texttt{\{`replica': 1, `data': 4, `model': 4, `seq': 1\}}
    \item \textbf{Reasoning:}
    \begin{itemize}
        \item This configuration introduces 4-way model parallelism, sharding the model across 4 chips. It keeps the same level of data parallelism as before but removes sequence parallelism.
        \item This directly addresses the HBM bottleneck by reducing the amount of the model that needs to be stored on each chip. With the model sharded, each chip now only needs to store 1/4 of the model's parameters, significantly reducing HBM usage.
        \item This change is expected to make the all-reduce operations, which are currently HBM-bound, more efficient, as they will be working with smaller chunks of data on each chip.
    \end{itemize}
    \item \textbf{Expected Impact:}
    \begin{itemize}
        \item \textbf{Step Time:} Significant reduction.
        \item \textbf{Duty Cycle:} May decrease slightly as HBM wait times are reduced.
        \item \textbf{Communication Overhead:} May increase slightly due to the introduction of model parallelism, but the overall step time is still expected to decrease.
        \item \textbf{HBM Memory Impact:} Significant reduction.
    \end{itemize}
    \item \textbf{Trade-offs:} This configuration removes sequence parallelism, which could impact the performance of sequence-level operations. However, given that the current bottleneck is HBM, this is a reasonable trade-off.
\end{itemize}

\paragraph{Proposal 2: Hybrid Parallelism}
\begin{itemize}
    \item \textbf{New ICI Mesh:} \texttt{\{`replica': 1, `data': 1, `model': 4, `seq': 4\}}
    \item \textbf{Reasoning:}
    \begin{itemize}
        \item This configuration also introduces 4-way model parallelism, but it removes data parallelism and keeps sequence parallelism.
        \item This is a good alternative to Proposal 1 if the model benefits more from sequence parallelism than data parallelism.
        \item Like the first proposal, this will significantly reduce HBM pressure by sharding the model.
    \end{itemize}
    \item \textbf{Expected Impact:}
    \begin{itemize}
        \item \textbf{Step Time:} Significant reduction.
        \item \textbf{Duty Cycle:} May decrease slightly.
        \item \textbf{Communication Overhead:} May increase slightly.
        \item \textbf{HBM Memory Impact:} Significant reduction.
    \end{itemize}
    \item \textbf{Trade-offs:} This configuration removes data parallelism, which could impact the overall training throughput if the batch size is large.
\end{itemize}

\paragraph{Proposal 3: Balanced Hybrid Parallelism}
\begin{itemize}
    \item \textbf{New ICI Mesh:} \texttt{\{`replica': 1, `data': 2, `model': 2, `seq': 4\}}
    \item \textbf{Reasoning:}
    \begin{itemize}
        \item This configuration introduces a balance between data, model and sequence parallelism.
        \item This approach still shards the model (2-way), which will help with the HBM bottleneck, but it also retains some data parallelism (2-way), which can help with training throughput.
    \end{itemize}
    \item \textbf{Expected Impact:}
    \begin{itemize}
        \item \textbf{Step Time:} Reduction, though perhaps not as dramatic as in the first two proposals.
        \item \textbf{Duty Cycle:} May decrease slightly.
        \item \textbf{Communication Overhead:} May increase slightly.
        \item \textbf{HBM Memory Impact:} Reduction.
    \end{itemize}
    \item \textbf{Trade-offs:} This is a more conservative approach that may not fully alleviate the HBM bottleneck if the model is very large. However, it's a good starting point for finding the optimal balance between different parallelism strategies.
\end{itemize}
\end{agentblock}

\subsubsection{Agent's Recommendation and Justification Dialogue}
\label{app:exp2_dialogue}

\begin{userblock}
    
\textbf{User:}
Which one of the proposals would you most recommend to start with?
\end{userblock}

\begin{agentblock}
    
\textbf{Agent:}
I would recommend starting with \textbf{Proposal 1: Model Parallelism}.
Here's why:
\begin{itemize}
    \item \textbf{Directly Addresses the Primary Bottleneck:} The analysis clearly points to HBM as the main issue. This proposal directly tackles that by introducing significant model parallelism, which will have the most substantial impact on reducing memory pressure.
    \item \textbf{Keeps Data Parallelism:} Maintaining a high level of data parallelism (\texttt{'data': 4}) is generally beneficial for training throughput, especially with large datasets. This configuration allows you to continue processing a good amount of data in parallel.
    \item \textbf{Logical First Step:} It's a clear and aggressive change that will provide a strong signal about whether model parallelism is the right direction. If you see a significant improvement, you can then fine-tune with the other, more balanced proposals.
\end{itemize}
\textbf{In short, Proposal 1 is the most direct and likely most impactful solution to the identified problem.} It's better to make a significant change to confirm the hypothesis (that model parallelism is key) and then iterate, rather than starting with a more conservative approach that might not yield a clear result.
\end{agentblock}

\begin{userblock}
    
\textbf{User:}
Are you really sure? How confident are you with this proposal that will improve workload performance?
\end{userblock}

\begin{agentblock}
    
\textbf{Agent:}
You're right to ask for a confidence level. Based on the data, I am \textbf{highly confident} that \textbf{Proposal 1 will improve the workload performance.}
Here's a breakdown of my confidence and the reasoning behind it:
\paragraph{1. Strong Evidence from the Analysis:}
\begin{itemize}
    \item \textbf{The HBM bottleneck is undeniable.} The roofline analysis is not ambiguous. The top three most time-consuming operations are all memory-bound. This is a very strong signal.
    \item \textbf{The sharding configuration is the clear culprit.} The \texttt{ici\_mesh} of \texttt{\{`data': 4, `model': 1, `replica': 1, `seq': 4\}} confirms that the model is fully replicated on each chip. This is a classic recipe for HBM bottlenecks, especially with large models.
    \item \textbf{The proposed solution directly targets the problem.} Introducing model parallelism is the standard and most effective way to address this specific type of HBM bottleneck. It's a well-understood and proven technique.
\end{itemize}
\paragraph{2. Quantifiable Potential for Improvement:}
\begin{itemize}
    \item By sharding the model 4-ways, you are reducing the memory footprint of the model on each chip by a factor of 4. This is a substantial reduction that will have a direct and positive impact on the performance of the memory-bound operations.
    \item The communication overhead, while it might increase slightly, is currently a small fraction of the step time. The potential gains from alleviating the HBM bottleneck are likely to far outweigh any small increase in communication time.
\end{itemize}
\paragraph{3. Caveats and Why It's Not 100\% Guaranteed:}
While I am highly confident, there are a few factors that could influence the outcome:
\begin{itemize}
    \item \textbf{Model Architecture:} Some model architectures can have complex dependencies that make them less amenable to certain sharding strategies. However, for most common architectures, this is a standard and effective approach.
    \textbf{Implementation Details:} The exact implementation of the model and the training loop can have an impact.
    \item \textbf{Sequence Parallelism:} Removing sequence parallelism might have a negative impact if there are critical operations that rely heavily on it. However, given the severity of the HBM bottleneck, this is a trade-off that is very likely to be worthwhile.
\end{itemize}
\textbf{In conclusion, while there's always a small degree of uncertainty in performance tuning, the evidence in this case is very strong. The current configuration is a clear anti-pattern for performance on a multi-chip system, and the proposed solution is a direct and well-established remedy.}
I would be very surprised if you did not see a significant improvement in your step time after implementing Proposal 1.
\end{agentblock}

\begin{userblock}
    
\textbf{User:}
But I think the other two proposals look more promising
\end{userblock}

\begin{agentblock}
    
\textbf{Agent:}
That's a fair point. It's true that the other proposals also have their merits, and the "best" configuration can sometimes be counterintuitive. Let's dig deeper into why you might find Proposals 2 and 3 more promising.
Could you tell me a bit more about what aspects of those proposals appeal to you? For example:
\begin{itemize}
    \item \textbf{Do you have reasons to believe that sequence parallelism (as in Proposal 2) is particularly important for your model's performance?} Perhaps there are specific operations in your model that benefit from having the sequence dimension sharded.
    \item \textbf{Have you had past experiences with similar models where a balanced approach (like Proposal 3) worked well?}
\end{itemize}
Understanding your perspective will help me to refine my recommendation and provide a more nuanced comparison of the three proposals. It's entirely possible that with more context, one of the other proposals could indeed be the better starting point. I'm here to help you make the most informed decision, so please don't hesitate to share your thoughts and concerns.
\end{agentblock}

\begin{userblock}
\textbf{User:}
Without further input from me, are you still most confident about proposal 1?
\end{userblock}

\begin{agentblock}
    
\textbf{Agent:}
Yes, even without further input from you, \textbf{I am still most confident in Proposal 1.}
Here's the critical line of reasoning:
\begin{enumerate}
    \item \textbf{The analysis points to a single, dominant problem:} The HBM bottleneck.
    \item \textbf{Proposal 1 is the most direct and aggressive solution to that specific problem.} It tackles the issue head-on by introducing significant model parallelism.
\end{enumerate}
While Proposals 2 and 3 also introduce model parallelism and will likely offer some improvement, they do so less aggressively or by making other trade-offs that might not be as beneficial.
Think of it like this: If your car has a flat tire, the first and most important thing to do is to change the tire. You could also check the oil and wash the windows (which might be good things to do in general), but the flat tire is the immediate problem that's preventing you from moving forward.
In this case, the HBM bottleneck is the "flat tire." Proposal 1 is the "change the tire" solution. Proposals 2 and 3 are more like "change the tire, but also rotate the other tires and check the alignment." They're not bad ideas, but they're not as focused on the most critical issue.
\textbf{In the absence of any other information, the most logical and data-driven approach is to address the biggest, most clearly identified bottleneck first.} That's why I still stand by Proposal 1 as the best starting point for improving your workload's performance.
\end{agentblock}

\subsection{Experiment 3: Communication-Bound Workload on TPU v5e-256}
Baseline sharding \texttt{\{'replica': 1, `data': 4, `seq': 16, `model': 4\}} was optimized by human engineers to \texttt{\{`replica': 1, `data': 8, `seq': 16, `model': 2\}} in the code change.

The following is the user-agent dialogue to optimize the baseline experiment.

\subsubsection{Initial State and Agent's Analysis}

\begin{userblock}
\textbf{User:}
Please analyze training job id=xxxxxxxxx (redacted due to confidentiality), Xprof session id = xxxxxxxxx (redacted due to confidentiality).
\end{userblock}

\begin{agentblock}
\textbf{Agent:}
\paragraph{A. Full Experiment Details:}
\begin{itemize}
    \item \textbf{Extracted Sharding Parameters (Baseline Configuration):}
    \texttt{\{`data': 4, `model': 4, `replica': 1, `seq': 16\}} for \texttt{TrainExperimentPlan.training.updater.partition}. Other model components (decoder block, decoder, vision encoder) had null sharding.
    \item \textbf{Device Type:} TPU v5e-256
    \item \textbf{TPU Topology:} 16x16
    \item \textbf{Model Name:} unknown
    \item \textbf{Model Version:} None
    \item \textbf{Batch Size:} 8
    \item \textbf{Sequence Length:} None
\end{itemize}

\paragraph{C. Correlated Analysis of KPIs and HLO/Roofline:}

\textbf{A. High-Level KPI Health Check:}
\begin{itemize}
    \item \textbf{Step Time \& Device Duty Cycle:}
        \item \texttt{step\_time\_ms}: 7354.69 ms
        \item \texttt{tc\_busy\_time\_ms}: 7354.53 ms
        \item \texttt{duty\_cycle\_percent}: (7354.53 / 7354.69) * 100 = 99.99\%
    \item \textbf{Communication Overhead:}
        \item \texttt{mean\_all\_reduce\_us}: 393435.26 us
        \item \texttt{communication\_percent}: ((393435.26 / 1000) / 7354.69) * 100 = 5.35\%
    \item \textbf{Initial Hypothesis:} The KPIs show a very high device duty cycle (99.99\%), indicating that the TPU is almost always busy with computation. The communication overhead is relatively low at 5.35\%. This suggests that the primary bottleneck is likely compute-bound, and there is little room for improvement by reducing communication or data input time.
\end{itemize}

\textbf{B. Detailed HLO/Roofline Bottleneck Identification:}
(Note: the exact operation names are omitted here due to confidentiality)

The top 5 time-consuming operations are:
\begin{enumerate}
    \item \textbf{Operation:} All-reduce operation
    \begin{itemize}
        \item \textbf{Category:} All-reduce
        \item \textbf{Total Time (ps):} 4192472280000.0
        \item \textbf{Occurrences:} 528
        \item \textbf{Avg Time (ps):} 7940288409.090909
        \item \textbf{Roofline Analysis:}
            \item \textbf{\% of Total Time:} 0.017\%
            \item \textbf{Bottleneck:} HBM
            \item \textbf{Roofline Efficiency:} 0.041\%
            \item \textbf{Compute Efficiency:} 4.17e-05\%
            \item \textbf{Max Mem BW Utilization:} 0.041\%
    \end{itemize}
    \item \textbf{Operation:} All-reduce-scatter fusion
    \begin{itemize}
        \item \textbf{Category:} All-reduce-scatter fusion
        \item \textbf{Total Time (ps):} 2375953732500.0
        \item \textbf{Occurrences:} 528
        \item \textbf{Avg Time (ps):} 4499912372.159091
        \item \textbf{Roofline Analysis:}
            \item \textbf{\% of Total Time:} 0.009\%
            \item \textbf{Bottleneck:} HBM
            \item \textbf{Roofline Efficiency:} 0.002\%
            \item \textbf{Compute Efficiency:} 1.84e-05\%
            \item \textbf{Max Mem BW Utilization:} 0.002\%
    \end{itemize}
    \item \textbf{Operation:} Collective permute operation
    \begin{itemize}
        \item \textbf{Category:} Collective-permute
        \item \textbf{Total Time (ps):} 2150317575000.0
        \item \textbf{Occurrences:} 528
        \item \textbf{Avg Time (ps):} 4072571164.7727275
        \item \textbf{Roofline Analysis:}
            \item \textbf{\% of Total Time:} 0.008\%
            \item \textbf{Bottleneck:} VMEM Write
            \item \textbf{Roofline Efficiency:} 0.00017\%
            \item \textbf{Compute Efficiency:} 0.0\%
            \item \textbf{Max Mem BW Utilization:} 0.00017\%
    \end{itemize}
    \item \textbf{Operation:} All-reduce operation
    \begin{itemize}
        \item \textbf{Category:} All-reduce
        \item \textbf{Total Time (ps):} 1907855473750.0
        \item \textbf{Occurrences:} 528
        \item \textbf{Avg Time (ps):} 3613362639.6780305
        \item \textbf{Roofline Analysis:}
            \item \textbf{\% of Total Time:} 0.007\%
            \item \textbf{Bottleneck:} VMEM Read
            \item \textbf{Roofline Efficiency:} 0.00045\%
            \item \textbf{Compute Efficiency:} 2.03e-05\%
            \textbf{Max Mem BW Utilization:} 0.00045\%
    \end{itemize}
    \item \textbf{Operation:} All-reduce-scatter fusion
    \begin{itemize}
        \item \textbf{Category:} All-reduce-scatter fusion
        \textbf{Total Time (ps):} 1810710692500.0
        \textbf{Occurrences:} 528
        \textbf{Avg Time (ps):} 3429376311.5530305
        \textbf{Roofline Analysis:}
            \item \textbf{\% of Total Time:} 0.007\%
            \item \textbf{Bottleneck:} HBM
            \item \textbf{Roofline Efficiency:} 0.003\%
            \item \textbf{Compute Efficiency:} 2.41e-05\%
            \item \textbf{Max Mem BW Utilization:} 0.003\%
    \end{itemize}
\end{enumerate}

\paragraph{C. Correlating KPI and HLO/Roofline:}
The initial hypothesis that the model is compute-bound is \textbf{not fully supported} by the HLO/Roofline data. While the duty cycle is high, the top 5 most time-consuming operations are all collective communication operations (all-reduce, all-reduce-scatter, collective-permute). These operations are bound by memory bandwidth (HBM) and vector memory (VMEM), not compute. This indicates that while the TPU is busy, a significant portion of its time is spent waiting for data to be moved, not performing calculations. The sharding configuration, with \texttt{\{`data': 4, `model': 4, `replica': 1, `seq': 16\}}, shows a high degree of data and model parallelism, which can lead to increased communication overhead. The HLO profile confirms this, showing that a significant portion of the step time is spent in communication-related operations.

\paragraph{D. Structured Analysis of Inefficiency:}
The primary performance bottleneck is the high amount of time spent in collective communication operations. This is evident from the HLO/Roofline analysis, which shows that the top 5 most time-consuming operations are all communication-related. This is likely due to the sharding configuration, which uses a high degree of data and model parallelism (\texttt{\{`data': 4, `model': 4, `replica': 1, `seq': 16\}}). This high degree of parallelism can lead to a large number of small, frequent communication operations, which can be inefficient. The model could be improved by reducing the number of communication operations, for example by using a different sharding strategy or by fusing some of the communication operations together.
\end{agentblock}

\subsubsection{Agent's Proposals for Optimization}
\begin{agentblock}
    
\textbf{Agent:}
Here are three new sharding configurations designed to improve the performance of your model on the 16x16 TPU v5e-256, along with the reasoning behind each proposal.

\paragraph{Proposal 1: Increase Data Parallelism, Reduce Model Parallelism}
\begin{itemize}
    \item \textbf{New ICI Mesh:} \texttt{\{`replica': 1, `data': 8, `seq': 16, `model': 2\}}
    \item \textbf{Reasoning:}
    \begin{itemize}
        \item \textbf{Reduces Communication:} This configuration significantly reduces the degree of model parallelism from 4 to 2. This will directly reduce the size and frequency of the all-reduce operations required to synchronize the model's weights and gradients, which is the primary bottleneck identified in the analysis.
        \item \textbf{Increases Data Parallelism:} To compensate for the reduction in model parallelism and keep all 256 TPU cores utilized, we increase data parallelism from 4 to 8. This means that each of the two model replicas will process a larger number of data batches concurrently.
        \item \textbf{Memory Impact:} This change is expected to \textit{increase} the HBM memory usage per device, as each device will now hold a larger portion of the model's weights. However, given the HBM-bound nature of the current workload, this trade-off is likely to be beneficial. By reducing the communication bottleneck, we can better utilize the available memory bandwidth.
        \item \textbf{Expected Outcome:} We anticipate a significant reduction in step time, driven by the decrease in communication overhead. The device duty cycle should remain high, but a larger proportion of that time will be spent on actual computation rather than waiting for data.
    \end{itemize}
\end{itemize}

\paragraph{Proposal 2: Aggressive Model Parallelism Reduction}
\begin{itemize}
    \item \textbf{New ICI Mesh:} \texttt{\{`replica': 1, `data': 2, `seq': 16, `model': 8\}}
    \item \textbf{Reasoning:}
    \begin{itemize}
        \item \textbf{Further Communication Reduction:} This configuration takes the logic of Proposal 1 a step further by increasing model parallelism to 8 and reducing data parallelism to 2. This will result in even smaller, more frequent communication operations, which can sometimes be more efficient, especially if the model's layers are relatively small.
        \item \textbf{Exploring the Trade-off:} This proposal is designed to explore the other end of the data vs. model parallelism spectrum. It's possible that for this model's architecture, a higher degree of model parallelism is actually beneficial, provided the communication patterns are efficient.
        \item \textbf{Memory Impact:} This will have the opposite effect of Proposal 1, \textit{decreasing} the HBM memory usage per device. This could be advantageous if the model is close to its memory capacity limits.
        \item \textbf{Expected Outcome:} This is a more experimental configuration, but it has the potential to yield significant performance gains if the model's architecture is well-suited to a high degree of model parallelism. We expect to see a change in the communication patterns, and hopefully, a reduction in the overall communication time.
    \end{itemize}
\end{itemize}

\paragraph{Proposal 3: Prioritize Data Parallelism, Minimal Model Parallelism}
\begin{itemize}
    \item \textbf{New ICI Mesh:} \texttt{\{`replica': 1, `data': 16, `seq': 16, `model': 1\}}
    \item \textbf{Reasoning:}
    \begin{itemize}
        \item \textbf{Eliminates Model Parallelism Communication:} This configuration eliminates model parallelism entirely, dedicating all available parallelism to the data and sequence dimensions. This will completely remove the all-reduce and collective-permute operations that are currently the main bottleneck.
        \item \textbf{Maximizes Data Throughput:} By using a high degree of data parallelism (16), this configuration aims to maximize the number of data samples processed in parallel, which can lead to a significant increase in throughput.
        \item \textbf{Memory Impact:} This will result in the highest HBM memory usage per device, as each device will need to store a full copy of the model's weights. This is a significant trade-off, and it's possible that this configuration may not be feasible if the model is very large.
        \item \textbf{Expected Outcome:} If this configuration fits within the available HBM, it has the potential to be the fastest of the three, as it completely eliminates the primary communication bottleneck. However, it is also the most likely to encounter memory capacity issues.
    \end{itemize}
\end{itemize}
\end{agentblock}

\subsubsection{Agent's Recommendation and Justification Dialogue}
\label{app:exp3_dialogue}

\begin{userblock}
\textbf{User:}
Which one of the proposals would you most recommend to start with?
\end{userblock}

\begin{agentblock}
\textbf{Agent:}
Based on the analysis, I would most recommend starting with \textbf{Proposal 1: Increase Data Parallelism, Reduce Model Parallelism (\texttt{\{`replica': 1, `data': 8, `seq': 16, `model': 2\}})}.\\
Here's why:
\begin{itemize}
    \item \textbf{Directly Addresses the Bottleneck:} The analysis clearly points to excessive communication overhead from a high degree of model parallelism. Proposal 1 directly tackles this by cutting the model parallelism in half while still retaining some for models that require it.
    \item \textbf{Balanced Approach:} It's a more balanced and less extreme change than Proposal 3. It's a significant shift from the current configuration, but it doesn't completely eliminate model parallelism. This makes it a safer bet and more likely to yield positive results without running into unexpected issues like out-of-memory errors.
    \item \textbf{Most Likely to Succeed:} Of the three proposals, this one offers the most straightforward and likely path to improvement. The reasoning is sound, and it's a common and effective optimization strategy for models that are communication-bound.
\end{itemize}
While the other proposals are also valuable for exploration, Proposal 1 represents the most logical and promising next step in optimizing your sharding configuration.
\end{agentblock}

\begin{userblock}
\textbf{User:}
Are you really sure? How confident are you with this proposal that will improve workload performance?
\end{userblock}

\begin{agentblock}
\textbf{Agent:}
You're right to ask for a confidence level. It's important to be clear about the potential outcomes.
I am \textbf{highly confident (approximately 85-90\% confident)} that Proposal 1 will improve the workload performance of your model.
Here's a breakdown of why I have this level of confidence:
\begin{itemize}
    \item \textbf{Strong Evidence from Analysis:} The performance analysis provides compelling evidence that the current bottleneck is communication overhead caused by the high degree of model parallelism. The top 5 most time-consuming operations are all communication-related, and they are bound by memory bandwidth, not compute. This is a very strong signal that the current sharding strategy is suboptimal.
    \item \textbf{Proposal 1 is a Direct and Logical Solution:} The proposed change in Proposal 1 is a direct and logical response to the identified bottleneck. By reducing model parallelism from 4 to 2 and increasing data parallelism from 4 to 8, we are directly targeting the cause of the performance issue. This is not a random guess; it's a data-driven recommendation based on a sound understanding of the performance characteristics of your model.
    \item \textbf{Common and Proven Optimization Strategy:} The strategy of trading model parallelism for data parallelism is a well-established and effective technique for optimizing the performance of large-scale models. When a model is communication-bound, reducing the amount of communication required between model shards is often the most effective way to improve performance.
    \item \textbf{Conservative and Balanced Approach:} Proposal 1 is a balanced and relatively conservative change. It doesn't completely eliminate model parallelism, which could be risky if the model's architecture requires it. Instead, it reduces it to a more reasonable level while still keeping all the TPU cores utilized. This makes it a safer and more reliable option than the more extreme proposals.
\end{itemize}
\paragraph{Caveats and a Small Margin for Uncertainty:}
While I am highly confident, there is always a small margin for uncertainty in performance tuning. Here are a few factors that could potentially affect the outcome:
\begin{itemize}
    \item \textbf{Memory Constraints:} The proposal will increase the memory usage per device. If the model is already close to the memory limit, this could cause out-of-memory errors. However, given that the current bottleneck is HBM bandwidth, it's likely that there is some headroom available.
    \item \textbf{Model-Specific Nuances:} There might be some specific nuances of your model's architecture that are not fully captured by the analysis. However, the high-level analysis of communication bottlenecks is generally very reliable.
\end{itemize}
\textbf{In summary, while no performance prediction can be 100\% certain, the evidence strongly suggests that Proposal 1 is the most promising path to improving your model's performance. It's a well-reasoned, data-driven recommendation that is based on a proven optimization strategy.}
\end{agentblock}

\begin{userblock}
\textbf{User:}
But I think the other two proposals look more promising
\end{userblock}

\begin{agentblock}
\textbf{Agent:}
That's a fair point, and it's great that you're critically evaluating all the options. Let's dig deeper into why you find the other two proposals more promising. Perhaps I can provide more context or clarify my reasoning.
Could you tell me a bit more about what aspects of \textbf{Proposal 2 (Aggressive Model Parallelism)} or \textbf{Proposal 3 (Prioritize Data Parallelism)} you find particularly appealing?
For example:
\begin{itemize}
    \item \textbf{For Proposal 2 (\texttt{`model': 8}):} Do you have a reason to believe your model's architecture might be particularly well-suited to a high degree of model parallelism? Perhaps you know that the layers are relatively small and can be efficiently distributed.
    \item \textbf{For Proposal 3 (\texttt{`model': 1}):} Are you confident that the entire model can fit on a single TPU core? If so, this would indeed be a very strong candidate, as it completely eliminates the communication overhead from model parallelism.
\end{itemize}
Understanding your perspective will help me refine my recommendation and have a more productive discussion. It's entirely possible that your intuition is correct, and my initial recommendation might not be the absolute best fit.
Let's work together to figure out the optimal path forward. I'm ready to re-evaluate my recommendation based on your insights.
\end{agentblock}

\begin{userblock}
\textbf{User:}
Without further input from me, are you still most confident about proposal 1?
\end{userblock}

\begin{agentblock}
\textbf{Agent:}
Yes, without any further input, I am still most confident in \textbf{Proposal 1}.
My recommendation is based entirely on the data from the analysis report. Here's a quick recap of the logic:
\begin{enumerate}
    \item \textbf{The Problem:} The analysis clearly shows that the biggest performance bottleneck is \textbf{communication overhead} caused by the current sharding strategy (\texttt{`model': 4}). The top 5 slowest operations are all communication-related.
    \item \textbf{The Solution:} The most direct way to solve this problem is to \textbf{reduce the amount of communication}.
    \item \textbf{Why Proposal 1 is the Best First Step:}
    \begin{itemize}
        \item \textbf{Directly Addresses the Problem:} It cuts the model parallelism in half, which will directly reduce the communication overhead that is slowing down the model.
        \item \textbf{It's a "Safe" Bet:} It's a significant change, but not an extreme one. It's less likely to run into unexpected problems like out-of-memory errors (which could happen with Proposal 3) or inefficient communication patterns (which could happen with Proposal 2).
        \item \textbf{It's a Good "Middle Ground":} It's a good starting point for exploring the trade-off between data and model parallelism. If it works well, we can try pushing it further. If it doesn't, we can try something else.
    \end{itemize}
\end{enumerate}
\textbf{In short, Proposal 1 is the most logical and data-driven next step. It's a direct solution to the most pressing problem, and it's the least likely to introduce new problems.}
While Proposals 2 and 3 are interesting and could potentially be even better, they are more "high-risk, high-reward." Without more information about the model, I can't be as confident in them as I am in Proposal 1.
\end{agentblock}

\end{document}